\documentclass[structabstract]{aa} 
\usepackage{graphicx}
\usepackage{txfonts}
\usepackage{url}
\usepackage{float}
\DeclareMathSymbol{\lesssim}{\mathrel}{AMSa}{"2E}
\begin{document}
\bibliographystyle{aa}
 \title{Optical and infrared polarimetry of the transient LMXB Cen X-4 in quiescence\thanks{Based on observations made with ESO Telescopes at the La Silla Observatory under programme ID 080.D-0849(A)}}

   \author{M. C. Baglio 
          \inst{1, 2, 3}          
          ,\         
          P. D'Avanzo \inst{1}
           ,\
          S. Campana \inst{1}
          \and
          S. Covino \inst{1}
          }
		  
   \institute{INAF, Osservatorio Astronomico di Brera, via E. Bianchi 46, I-23807 Merate (Lc), Italy\\
              \email{cristina.baglio@brera.inaf.it}
         \
            \\ 
            \and
            Universit\`{a} degli Studi di Milano, Dipartimento di Fisica, Via Celoria 16, I-20133 Milano, Italy
             \\
             \and 
             Universit\`{a} dell'Insubria, Dipartimento di Fisica, Via Valleggio 11, I–22100     Como, Italy 
             }             

   \date{ }

   \abstract
   {}
   {We present the first optical and infrared polarimetric study of the low mass transient X-ray binary Centaurus X-4 during its quiescent phase. This work is aimed to search for an intrinsic linear polarisation component in the system emitted radiation that might be due, e.g., to synchrotron emission from a compact jet, or to Thomson scattering with free electrons in an accretion disc. }
   {Multiband ($ BVRI $) optical polarimetric observations were obtained during two nights in 2008 at the ESO La Silla 3.6 m telescope, equipped with the EFOSC2 instrument used in polarimetric mode. These observations cover about the 30$ \% $ of the 15.1 hours orbital period. $ J $-band observations were obtained in 2007 with the NICS instrument on the TNG telescope at La Palma, for a totality of 1 hour observation.}
   {We obtained 3$ \sigma $ upper limits to the polarisation degree in all the optical bands, with the most constraining one being in the $ I $-band, where $ P_{\rm I}<0.5\% $. No significant phase-correlated variability has been noticed in all the filters. The $ J $-band observations provided a 6$ \% $ upper limit on the polarisation level.}
   {The constraining upper limits to the polarisation in the optical (above all in the $ I $-band), allowed us to evaluate the contribution of the possible emission of a relativistic particles jet to the total system radiation to be less then the 10$ \% $.This is in agreement with the observation of a spectral energy distribution typical of a single black body of a K-spectral type main sequence star irradiated from the compact object, without any significant additional component in the infrared. Due to the low S/N ratio it was not possible to investigate the possible dependency of the polarisation degree from the wavelength, that could be suggestive of polarisation induced by Thomson scattering of radiation with free electrons in the outer part of the accretion disc. Observations with higher S/N ratio are required to examine in depth this hypothesis, searching for significant phase-correlated variability.}

   \keywords{
               }
\authorrunning{Baglio, D'Avanzo, Campana \& Covino} 
\titlerunning{Optical and infrared polarimetry of the transient LMXB Centaurus X-4 in quiescence}
\maketitle

\section{Introduction}
Only a few low-mass X-ray binaries (LMXBs -- systems where a compact object like a neutron star or a black hole accretes matter from a companion star via Roche lobe overflow) have been studied with polarimetric techniques to date (Charles et al. 1980; Dolan \& Tapia 1989; Gliozzi et al. 1989; Hannikainen et al. 2000; Shultz et al. 2004; Brocksopp et al. 2007; Shahbaz et al. 2008; Russell et al. 2008; Russell et al. 2011). Polarisation provides a powerful diagnostic tool to obtain information about geometrical and physical conditions of these systems, scattering properties of their accretion discs or the presence of strong magnetic fields.

Most of the LMXB radiation is expected to be unpolarized. Optical light from LMXBs is in fact principally made of thermal blackbody radiation from accretion disc and the companion star, and does not possess any preferential direction of oscillation.
Hydrogen in the disc is nevertheless in many cases totally ionised; for this reason a significant (but small) linear polarisation (LP) is expected in the optical (Dolan 1984; Cheng et al. 1988) due to Thomson scattering of emitted unpolarised radiation with free electrons in the disc. This linear polarisation component is usually almost constant for scattering in the nearly symmetrical accretion disc. If there are deviations from axial symmetry, some phase-dependent variations might be expected. Furthermore, radiation emitted from the accretion disc could interact via inverse Compton scattering with the electrons in a hot plasma corona that surrounds the disc itself (Haardt et al. 1993). This phenomenon could induce high frequency polarisation.

Another possible and intriguing origin of a significant polarisation can be synchrotron emission, that arises from emission of a relativistic particle jet. Optically thin synchrotron radiation produces in fact intrinsically linearly polarized light at a high level, up to tens per cent, especially in the NIR (Russell \& Fender 2008). Jets in X-ray binaries are expected to be linked to accretion (disc-jet coupling, Fender 2001b), and for this reason they have been principally observed in persistent systems or during the outbursts of transient LMXBs, especially if containing a black hole. In the past few years, the
evidence for jet emission during quiescence of LMXBs, containing both black holes and neutron stars have been reported (Russell et al. 2006; Russell et al. 2007; Russell \& Fender 2008; Baglio et al. 2013; Shahbaz et al. 2013). The detection of a high level of linear polarisation in the NIR is for this reason considered as the main route to assess for the emission of a relativistic jet.

Radiation from any source can also be polarised by the interaction with interstellar dust. This effect depends on wavelength as described by the Serkowski law Serkowski et al. 1975 and must be accounted for in the analysis.

Transient LMXBs are generally faint objects in the optical; for this reason only the brightest ones have been observed in polarimetry. The most part of these studies regarded systems during outbursts or systems with BHs as compact object during quiescence.  

In this paper we report the results of the optical multi-band ($ BVRI $) and infrared ($ J $-band) polarimetric observations of the LMXB Cen X-4 during quiescence using the ESO 3.6 m telescope at La Silla and the TNG telescope, respectively. This is the first polarimetric study of a quiescent LMXB containing a NS. 

Cen X-4 was discovered during an X-ray outburst in 1969 by the X-ray satellite \textit{Vela 5B} (Conner et al. 1969). During a second outburst in 1979 the source was detected also in the radio band (Hjellming 1979), and its optical counterpart was identified with a blue star that had brightened by 6 mag to $ V $=13 (Canizares et al. 1980). The companion star was at a later stage classified as a $ 0.7M_{\odot} $ K5--7 star (Shahbaz et al. 1993; Torres et al. 2002), that evolved in order to fill its $ \sim 0.6\,R_{\odot} $ Roche lobe. The $ \sim 15.1 $ hrs orbital period has been determined thanks to the optical light curve ellipsoidal variations (Cowley et al. 1988; Chevalier et al. 1989; McClintock et al. 1990). 
Cen X-4 is one of the brightest quiescent systems in the optical known to date ($V$=18.7 mag) and possesses a non-negligible disc component in the optical that contributes $ \sim 80 \% $ in $ B $, $ \sim 30 \% $ in $ V $, $ 25 \% $ in $ R $ and $ 10 \% $ in $ I $ (Shahbaz et al. 1993; Torres et al. 2002; D'avanzo et al. 2005). Cen X-4 is at a distance of $ 1.2 \pm 0.3 $ kpc (Kaluzienski et al. 1980) and the interstellar absorption is low ($ A_{V}=0.3 \,\rm mag $).
These characteristics make Cen X-4 an excellent candidate for polarimetric studies in quiescence.

Throughout the paper all the uncertainties are at $ 68 \% $ confidence level unless stated differently.
\section{Optical polarimetry}\label{Obs_parag}
The system Cen X-4 was observed on 11-12 March 2008 with the ESO 3.6 m telescope at La Silla, using the EFOSC2 instrument in polarimetric mode with the optical $ BVRI $ filters ($ 440 \rm nm -793 \rm nm $). The nights were clear, with seeing $ \lesssim 1'' $. 
Image reduction was carried out following standard procedures: subtraction of an averaged bias frame and division by a normalized regular flat frame. 
All the flux measures have been done thanks to accurate aperture photometry made with 
\textit{\tt daophot} (Stetson 1987) for
all the objects in the field. 
The polarimetric calibration was done against two polarimetric standard polarised and non-polarised stars provided by the FORS consortium based on Commissioning data taken with FORS1 \footnote{\url{www.eso.org/sci/facilities/paranal/instruments/fors/inst/pola.html}}.


A Wollaston prism has been inserted in the optical path. The incident radiation was split into two simultaneous and orthogonally polarised ordinary and extraordinary beams (o- and e- beams). Thanks to a Wollaston mask the different images do not overlap. The use of a rotating half wave plate (HWP) allowed us to take images at four different angles with respect to the telescope axis ($ \Phi_{i} = 22.5^{\circ}(i-1)$, $i=0,1,2,3$). Alternating the filters, a set of 10 images of 90 seconds integration were obtained for each HWP angle for the $ BVI $ filters and 9 for each angle of the $ R $- band, divided in the two nights of observations, covering about the $ 30 \% $ of the orbital period. 

The normalised Stokes parameters $ Q $ and $ U $ for linear polarisation (LP) of the observed radiation are commonly evaluated starting from flux measures in both o- and e- beams ($ f^{o} $, $ f^{e} $) at just two orientation angles of the telescope's axis ($ 0^{\circ} $ and $ 45^{\circ} $):

\begin{equation}
Q=\frac{f^{o}(0^{\circ})-f^{e}(0^{\circ})}{f^{o}(0^{\circ})+f^{e}(0^{\circ})} ; \,\,\,\,\, U=\frac{f^{o}(45^{\circ})-f^{e}(45^{\circ})}{f^{o}(45^{\circ})+f^{e}(45^{\circ})}.
\end{equation}

Thanks to the rotating HWP, a higher accuracy measure of $ Q $ and $ U $ has been possible, using the whole set of possible orientations. If $ \Phi_{i}$ are the HWP orientation angles defined as above, $ Q $ and $ U $ can be obtained from:

\begin{equation}\label{stokes}
Q=\frac{F(\Phi_{1})-F(\Phi_{3})}{2} ; \,\,\,\,\, U=\frac{F(\Phi_{2})-F(\Phi_{4})}{2},
\end{equation}
where
\begin{equation}
F(\Phi_{i})=\frac{f^{o}(\Phi_{i})-f^{e}(\Phi_{i})}{f^{o}(\Phi_{i})+f^{e}(\Phi_{i})}.
\end{equation}

A first raw estimate of the observed polarisation degree $ P_{\rm obs} $ and angle $ \theta $ can then be obtained as:
\begin{equation}\label{Pobs}
P_{\rm obs}=(U^{2}+Q^{2})^{0.5}
\end{equation}
\begin{equation}
\theta = 0.5 \tan^{-1} $(U/Q)$.
\end{equation}

Since the Stokes parameters statistics is not Gaussian (Wardle \& Kronberg 1974; di Serego Alighieri 1998) the calculated values of linear polarisation have to be corrected for a bias factor. The real polarisation degree $ P $ can be obtained from:
\begin{equation}\label{bias}
P=P_{\rm obs}\sqrt{1-\left(\frac{\sigma_{\rm P}}{P_{\rm obs}}\right)^{2} },
\end{equation}
where $ \sigma_{\rm P} $ is the r.m.s. error on the polarisation degree.

\subsection{Averaged Stokes parameters for LP}
 \begin{figure}
\begin{center}
\includegraphics[scale=0.25]{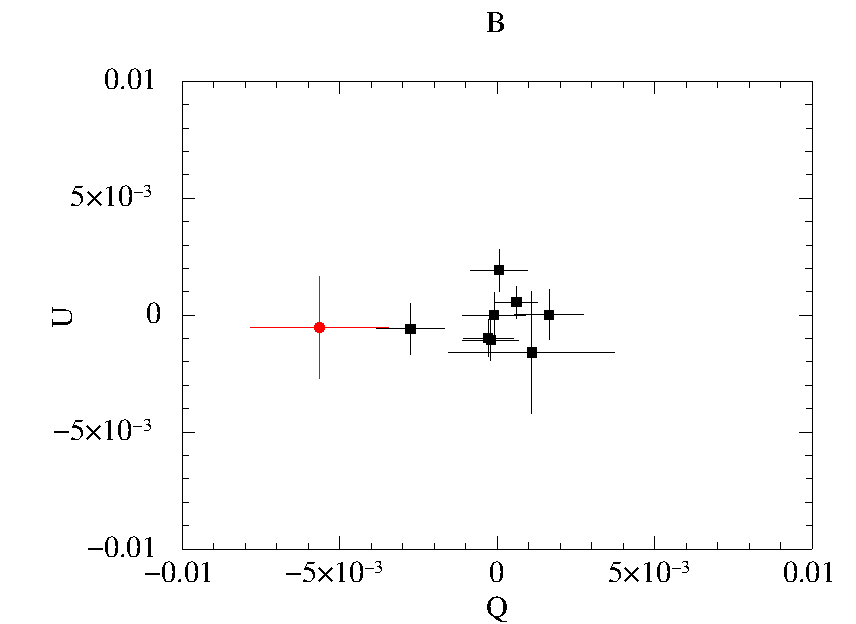}
\includegraphics[scale=0.25]{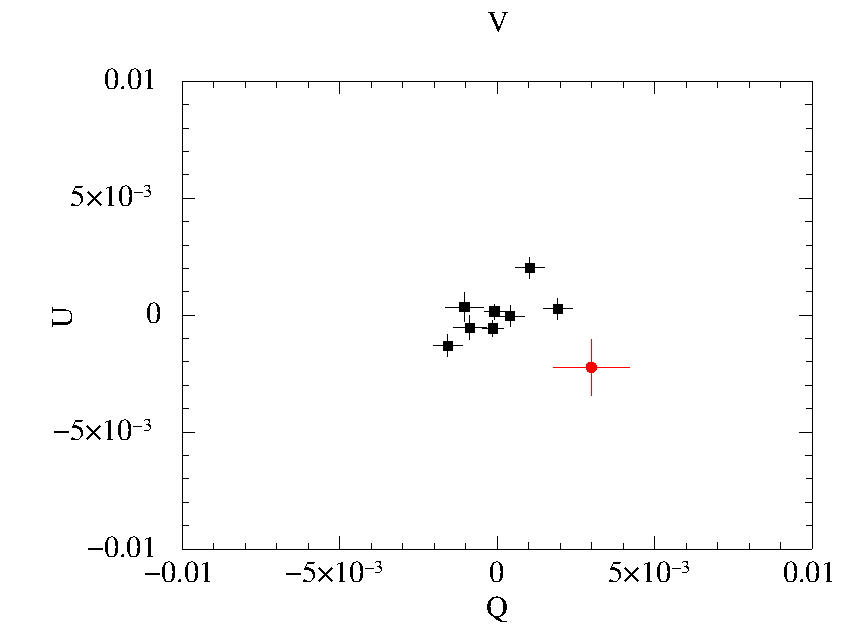}
\includegraphics[scale=0.25]{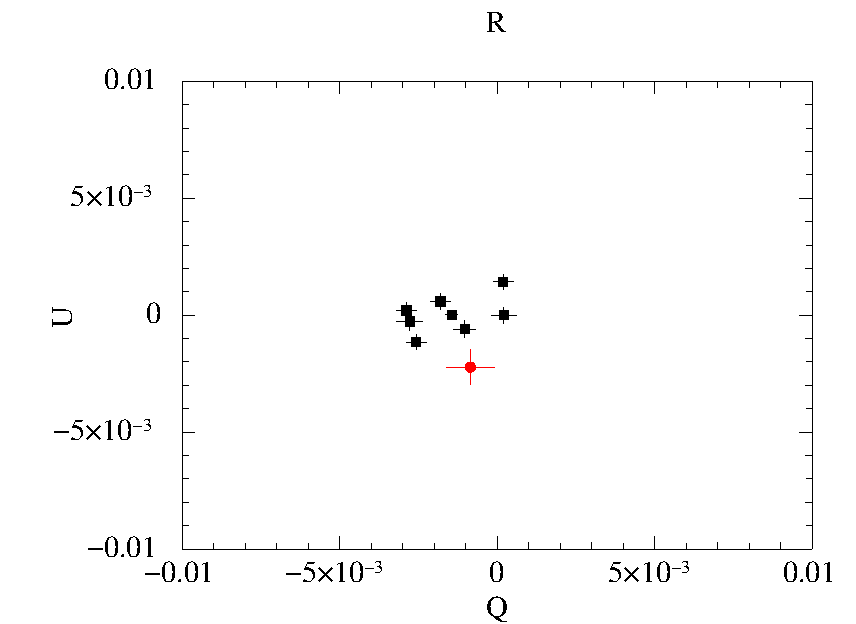}
\includegraphics[scale=0.25]{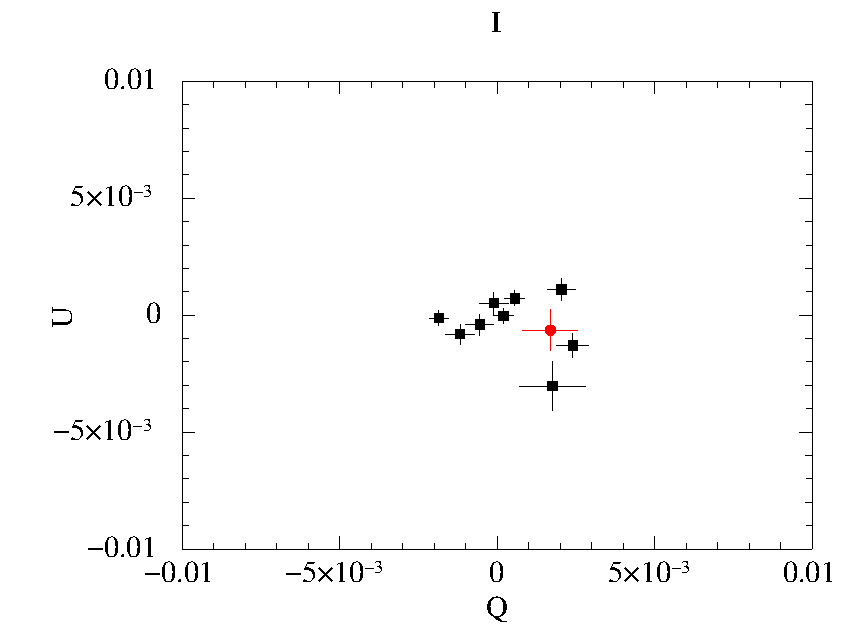}
\caption{$ U $ vs. $ Q $ for the averaged images of the four optical filters $ BVRI $, for Cen X-4 (red dots) and for eight reference field stars (black squares). The represented Stokes parameters have been already corrected for the effect of instrumental polarisation, as described in Sec. 2.1.}
\label{U_Q_medie}
\end{center}
\end{figure}
We obtained the averaged (over all observations) Stokes parameters $ Q $ and $ U $ (eq.\ref{stokes}) in all the filters analysed, for Cen X-4 and a selection of eight isolated and non-saturated reference stars, supposed to be unpolarised. For $ Q $ and $ U $ separately, a correction with respect to the 
average Stokes parameters measured for the reference field stars in each filter was then achieved. This correction results to be small (at most the $0.1\%$), supporting our hypothesis of non-polarisation of the reference stars, and is crucial in order to eliminate most of the instrumental contribution to the Stokes parameters of the target. The same correction has been applied to the reference stars themselves.
The errors on $ Q $ and $ U $ were found simply by propagating the photometric errors with the uncertainties obtained for the unpolarised field stars averaged Stokes parameters.

As shown in Fig. \ref{U_Q_medie}, the Stokes parameters evaluated for the reference stars and corrected as explained above form a cluster around 0 in each filter. The fact that the target might show a different behaviour with respect to this selection of unpolarised stars can be due to a higher interstellar absorption (if distance is larger than that of the stars), or to a real intrinsic polarisation component. This comparative method has been used for gamma-ray bursts afterglows (Covino et al. 1999) and for X-Ray Binaries (Dubus \& Chaty 2006).

The only filters for which Cen X-4 lies at a $ \gtrsim $3$ \sigma $ level from the weighted mean of the reference stars are the $ V $- and $ R $- bands, whereas in $ B $- and $ I $- bands the source's Stokes parameters are comparable with those of the reference stars. 
To evaluate the polarisation degree $ P $, one can use eq. \ref{Pobs}, with the correction reported in eq. \ref{bias}. This method is nevertheless not advised in case the r.m.s. error on $ P_{\rm obs} $ is comparable to $ P_{\rm obs} $ itself. In particular, in our case, the smallest obtained error bar for Cen X-4 corresponds to $ \sim 50 \% $ of the measure (except for the $ I $- band, where the S/N ratio is higher). In this case the best way to estimate $ P $ and the polarisation angle $ \theta $ is reported in the next section. 

\subsection{The $ S $- parameter}\label{S_paragraph}
The simple calculus proposed in eq. \ref{Pobs} does not take care of the possible interstellar polarisation and of Wollaston prism's imperfections. For this reason it is necessary to apply a correction to the Stokes parameters obtained with eq. \ref{stokes}, relying on polarimetric standard stars. Alternatively, it is possible to evaluate, for each HWP angle $ \Phi $ defined as in Sec. \ref{Obs_parag}, a parameter $ S\left(\Phi\right)  $ starting from the o- and e-fluxes of the target, $ f^{o}(\Phi) $ and $ f^{e}(\Phi) $, and from the averaged ratio between o- and e- fluxes of some unpolarised field stars $ f^{o}_{u}(\Phi) $ and $ f^{e}_{u}(\Phi) $. In particular (di Serego Alighieri 1998; Covino et al. 1999):
\begin{equation}
S(\Phi)=\left(\frac{f^{o}(\Phi)/f^{e}(\Phi)}{\left\langle f^{o}_{u}(\Phi)/f^{e}_{u}(\Phi_{i})\right\rangle  }-1\right)/\left(\frac{f^{o}(\Phi)/f^{e}(\Phi)}{\left\langle f^{o}_{u}(\Phi)/f^{e}_{u}(\Phi)\right\rangle  }+1\right).
\end{equation}

This parameter can be regarded as the component of the normalised Stokes vector that describes LP along the direction selected by the HWP angle $ \Phi $. The relation between $ S $, $ P $ and $ \theta $ is given by:
\begin{equation}\label{fit_cos}
S(\Phi)= P\cos 2\left( \theta -\Phi\right).
\end{equation}

\begin{figure}
\centering
\includegraphics[scale=0.25]{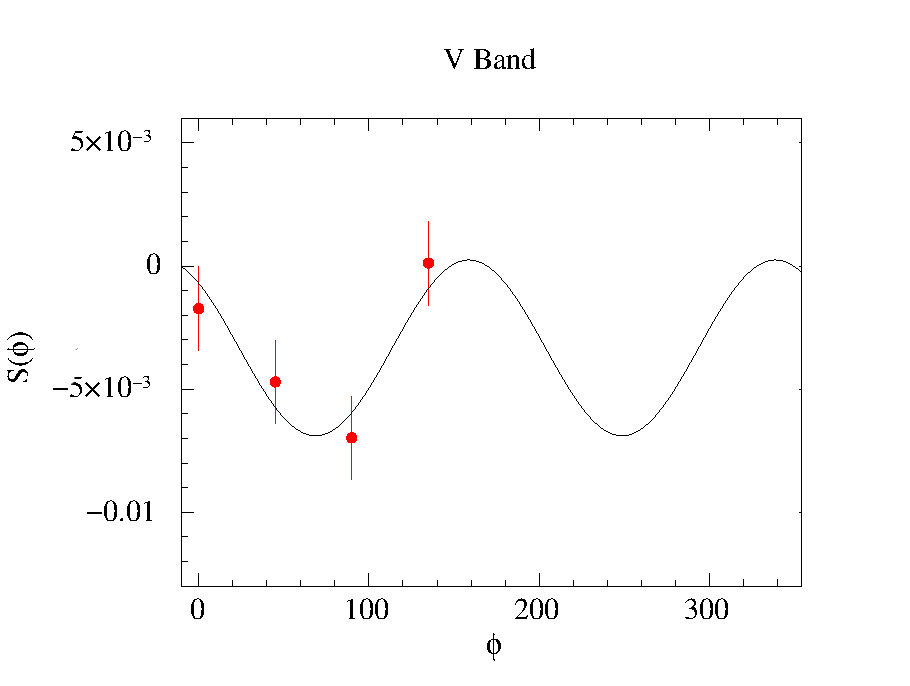}
\caption{Cosinusoidal $ V $-band fit of S$ (\Phi) $ for Cen X-4.}
\label{S_V_fit}
\end{figure}

In this way, one can fit the function $ S(\Phi) $ with eq. \ref{fit_cos} and obtain $ P $, $ \theta $ and their errors from the semi-amplitude of the curve and from the $x$-value that corresponds to the first curve's maximum, respectively. With this method the measure of a polarimetric standard star becomes useless, since the parameter $ S $ is already normalised to the non-polarised reference stars, that all cluster around the same point of the plane ($Q$,$U$) in each filter. In particular, the so-obtained values should be automatically corrected for interstellar and instrumental effects, and do not need any bias correction (eq. \ref{bias}). The cosinusoidal fits of $ S(\Phi) $ for the $ V- $band is reported in Fig. \ref{S_V_fit}.

Adopting this method we derived the polarisation levels of Cen X-4 in the four optical bands. Only in the $ V $ and $ R $-band there is a detection of polarisation different from 0 within 1$\sigma$.
 Since at least a 3$ \sigma $ over zero polarisation degree would be an evidence of an intrinsic polarisation detection, we decided to quote the 3$ \sigma $ upper limits in all bands (Tab. \ref{pola}). As reported in Tab. \ref{pola}, the 3$ \sigma $ upper limit for the $ I $ band is really tightly constraining, due to the higher measured S/N ratio with respect to the other bands.
No evidence of a significant wavelength dependency of P is observed, since the polarisation measures in all bands are consistent with each other within 2$ \sigma $.

\begin{table}
\caption{Values of the $ V $ and $ R $-band detected polarisation degrees $ P $, and $ BVRI $-band 3$ \sigma $ upper limits and polarisation angles for Cen X-4. The errors have been evaluated using contour plots in each band. In particular, in order to obtain the 68$\%$ c.l. we used $ \Delta \chi^{2} =2.3 $ as requested for fits with two free parameters.  }             
\label{pola}      
\centering                          
\begin{tabular}{|c c c c|}        
\hline\hline                 
$ B $ & $ V $ & $ R $ & $ I $\\    
\hline                        
\multicolumn{4}{|c|}{P ($\%  $)}\\ \hline
- & $ 0.36 \pm 0.18 $ & $ 0.19 \pm 0.16 $ & -\\
\hline
\multicolumn{4}{|c|}{P (3$\sigma$ upper limit)}                     \\ \hline
 $1.46 \%$ & $ 0.90\% $ & $ 0.67\% $ & $ 0.46\%  $ \\  
\hline                                   
\multicolumn{4}{|c|}{$ \theta $ ($ ^{\circ} $)}                               \\ \hline
 $ 159.41 ^{+49.59}_{-51.42} $ & $ 158.75 ^{+14.25}_{-15.75} $ & $ 167.81 ^{+24.19} _{-25.81}  $ & $ 189.87\pm 49.13$\\
\hline                                   
\end{tabular}
\end{table}

Following Serkowski et al. (1975) we were then able to evaluate the maximum expected interstellar contribution to the LP ($ P_{\rm max} $) of Cen X-4. In particular we retrieved the empirical formula $ P_{\rm max}\leq 3A_{V} $, according to which the maximum contribution to the LP for Cen X-4 due to interstellar effects should remain under the $ 0.9 \% $ level, consistently with our results. 

\subsection{Search for phase dependent variations of $ P $}
The values reported in Table \ref{pola} have been obtained by summing all the images taken for each filter; this increased the S/N ratio for our measure, but let us loose all the information about possible orbital phase-correlated variations of $ P $, $ Q $ and $ U $. In order to investigate the possible variability of the polarisation along $ P_{\rm orb} $, we analysed the single images calculating the orbital phases thanks to the precise ephemerides of Casares et al. (2007). Due to the large error bars however, the data proved inconclusive. 

\begin{figure}
\centering
\includegraphics[scale=0.25]{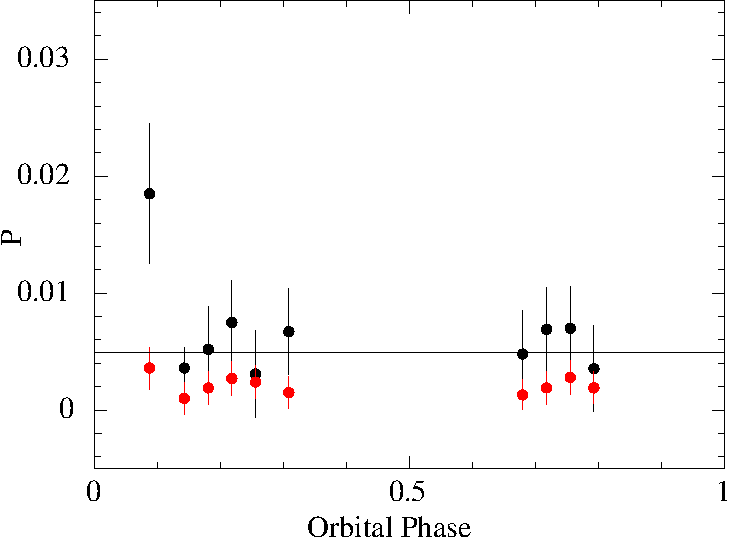}
\caption{$ V $-band polarisation curve of Cen X-4 (black squares) and of one field star (red dots), with superimposed the linear fit to the data.}
\label{V_phase_orb}
\end{figure}

The only variation from the linear fit to the data in all the analysed filters has been found during the first epoch in the $ V $-band (Fig. \ref{V_phase_orb}). In fact at phase 0.1 (that is, when the companion star contribution is near to its minimum) we measured a polarisation degree of $ 1.85\% \pm 0.60\% $, i.e. $ 2.3 \sigma$ from the average.

However, no similar features have been observed in the other bands, and furthermore this sudden increase in polarisation does not coincide with an increase or decrease of the observed flux. For these reasons we conclude to have observed an effect linked to photon statistics, and that any variability of polarisation intrinsic to Cen X-4 is too weak to be detected from our dataset, with the relatively low S/N ratio we have obtained in our observations.

\section{Infrared polarimetry}

The LMXB Cen X-4 was observed in IR ($ J $-band, 1.27$ \mu $m) on the 24$^{\rm th}$ April 2007 with the FGG TNG telescope at La Palma, equipped with the NICS instrument used in polarimetric mode. 
A set of 20 images of the field of 180 s integration each was obtained. A Wollaston prism was inserted in the grism wheel, that split incident radiation into the four polarisation components ($ 0^{\circ} $, $ 45^{\circ} $, $ 90^{\circ} $ and $ 135^{\circ} $ with respect to the telescope's axis), which were re-imaged at different positions along the Y-axis. 

Image reduction was carried out by subtracting the sky background to each frame.

Because of a technical problem with the instrument derotator, only the 25$ \% $ of the images have been used for our purposes. The selected images have been summed together in order to obtain a single image with an exposure time of 900 s.
Flux measures have been made thanks to aperture photometry performed with \textit{\tt daophot} for
all the objects in the field. 

The normalised Stokes parameters $ Q $ and $ U $ have been calculated from\footnote{\url{www.tng.iac.es/instruments/nics/files/pol_obs_v03.pdf}}
\begin{equation}
Q=\frac{f(0^{\circ})-f(90^{\circ})}{f(0^{\circ})+f(90^{\circ})}; \,\,\,\,\, U=\frac{f(45^{\circ})-f(135^{\circ})}{f(45^{\circ})+f(135^{\circ})},
\end{equation}
whereas the polarisation degree $ P $ can be obtained from eq. \ref{Pobs}.

\subsection{Results}
\begin{figure}
\centering
\includegraphics[scale=0.25]{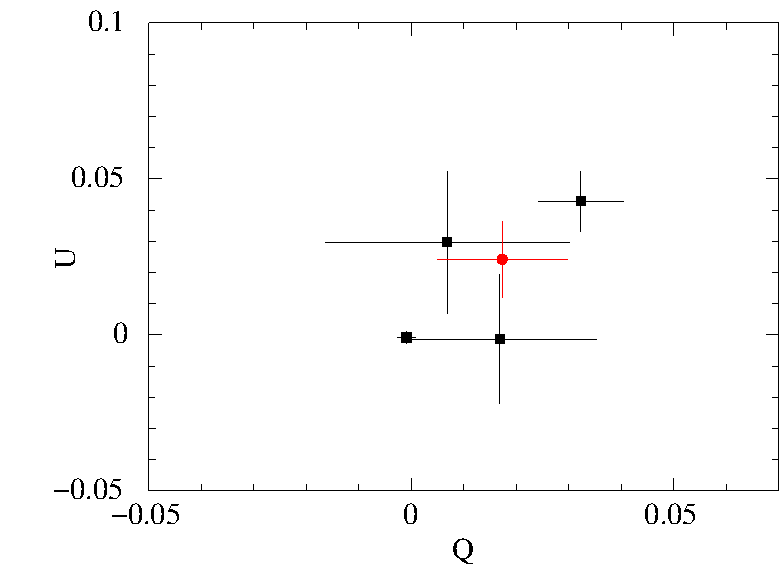}
\caption{U vs. Q for the $ J $-band observations, for Cen X-4 (red dot) and for four isolated reference stars (black squares). All the Stokes parameters values have been corrected for the weighted mean of $ Q $ and $ U $ of the field stars.}
\label{Q_U_J}
\end{figure}

We selected a sample of four isolated and supposed to be unpolarised stars in the field, in a region of the image as near as possible to the target, in order to make a comparison with Cen X-4, and we calculated $ Q $ and $ U $ for all of them. As we did for optical data, we represented the Stokes parameters in the $ Q-U $ plane and we observed that Cen X-4 was comparable to the field stars, that all cluster around a common value, corresponding to the possible instrumental polarisation contribution. In Fig. \ref{Q_U_J} we reported the $ Q $ vs. $ U $ plot for Cen X-4 and the four reference stars, with all the parameters corrected for the weighted mean of the field stars Stokes parameters.

Due to the large error bars measured for both the target and the field stars, we decided to evaluate directly a 3$ \sigma $ upper limit to the $ J $-band polarisation. We performed a Monte Carlo simulation starting from the values of $ Q $ and $ U $ of Cen X-4 corrected for the average $ Q $ and $ U $ measured for the field stars and we obtained an upper limit to $ P $ of $\sim 6 \% $ within $ 3\sigma $ (the upper limit is of $\sim 4 \% $ at a $ 1\sigma $ confidence level). 

\section{Discussion}
\subsection{Jets in quiescence?}
As stated in Russell et al. 2011, a polarimetric signature of the emission of a relativistic particles jet in LMXB is theoretically detectable both in the infrared and in the optical. In black holes and neutron stars X-ray binaries the linear polarisation degree of the emitted radiation is expected not to exceed the few $ \% $, with evidence of variability on short time-scales. This fact seems to suggest that a tangled and turbulent (no more than partially ordered) magnetic field is generally present at the base of the jet (there was a unique detection of a highly ordered magnetic field in a X-ray binary for the persistent system Cyg X-1, Russell \& Shahbaz 2014). Something similar seems to happen also in case of compact jets from active galactic nuclei (AGN), in particular in gigahertz peaked-spectrum sources, where the low level of polarisation ($1-7 \%$) measured in the optically thin regime is interpreted with the presence of a tangled magnetic field (O'Dea 1998). Also the flat-spectrum radio jets of AGN usually possess a low polarisation degree ($1-5 \%$), similarly to what happens to the radio jets of X-ray binaries, suggesting for tangled and helical magnetic fields (Helmboldt et al. 2007, Perlman et al. 2011). 

For Cen X-4, the measured degree of linear polarisation averaged over all observations is $\lesssim 1.5\% $ in the optical $ BVRI $ filters, and $\lesssim 6\% $ in the $ J $-band within 3$ \sigma $ (Table \ref{pola}). Under the hypothesis that neutron stars X-ray binaries could have a similar behaviour to that of black holes X-ray binaries in terms of jet emission, these upper limits are effectively consistent with the possible emission of a jet, with the low polarisation degrees possibly linked to the tangled and non-ordered structure of the magnetic field in a region close to the one where the jet is launched. However, no observational constraints of how tangled neutron star jets magnetic fields are exist to date. 

A possible way to search for the presence of the relativistic particle jet is to observe the spectral energy distribution (SED) of the system Cen X-4, as stated in Fender (2001a). In fact, in case of jet emission, an excess in the SED in the range of frequencies where the synchrotron emission gives its maximum contribution (NIR) is expected. The emission in this case would be characterised by an optically thick synchrotron radio spectrum (that corresponds to $ \alpha\geq 0 $, where $ \alpha $ is the spectral index and the flux density $ F_{\nu} $ is proportional to $ \nu^{\alpha} $), that should switch to an optically thin synchrotron spectrum at shorter wavelengths (i.e. $ \alpha \simeq -0.6 $). The break
frequency is expected to fall in the mid-Infrared (Fender 2001b; Corbel \& Fender 2002; Gallo et al. 2007; Migliari et al. 2007; Pe'er \& Casella 2009; Migliari et al. 2010; Rahoui et al. 2011; Gandhi et al. 2011). 

\begin{figure}
\centering
\includegraphics[scale=0.4]{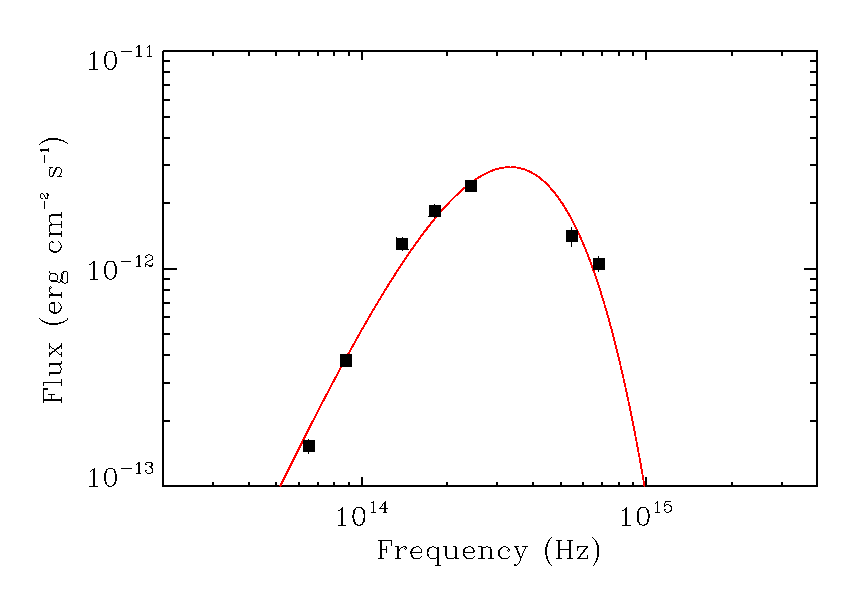}
\caption{Spectral energy distribution (SED) of Cen X-4 built starting from IR archival data and from the optical fluxes obtained by Cackett et al. (2013). The three datasets are not contemporary. Superimposed, the fit of the data with a black body of an irradiated star.}
\label{sed}
\end{figure}

We thus built the SED of the system (Fig. \ref{sed}), starting from IR data taken from the WISE All Sky and 2MASS catalogues\footnote{\url{irsa.ipac.caltech.edu/workspace/TMP_qLHgEi_32284/Gator/irsa/32533/tbview.html}} and from the optical $ V $- and $ B $-band data obtained from Cackett et al. (2013) during quiescence for Cen X-4. The fluxes values considered for the SED are reported in Tab. \ref{fluxes_tab}.

\begin{table}
\caption{$ IR $ and optical dereddened fluxes obtained from the WISE and 2MASS catalogues and from Cackett et al. (2013), used to build the multi-wavelength SED of Cen X-4 (Fig. \ref{sed}).    }             
\label{fluxes_tab}      
\centering                          
\begin{tabular}{|c c|}        
\hline\hline                 
Band & Flux ($\rm erg \cdot\rm cm^{-2}\cdot s^{-1}$)\\
\hline
 $ W_{\rm 2} $ WISE ($ \lambda = 4.6\, \mu \rm m $) & $ (1.53 \pm 0.12)\times 10^{-13} $  \\
 $ W_{\rm 1} $ WISE ($ \lambda = 3.4\, \mu \rm m $) & $ (3.79 \pm 0.13)\times 10^{-13} $  \\
 \hline
 $ K $ 2MASS ($ \lambda = 2.16\, \mu \rm m $) & $ (1.30 \pm 0.10)\times 10^{-12} $  \\
 $ H $ 2MASS ($ \lambda = 1.66\, \mu \rm m $) & $ (1.85 \pm 0.13)\times 10^{-12} $  \\
 $ J $ 2MASS ($ \lambda = 1.23\, \mu \rm m $) & $ (2.40 \pm 0.13)\times 10^{-12} $  \\
 \hline
 $ V $ (Cackett et al. 2013, $ \lambda = 5468\, \AA $) & $ (1.41 \pm 0.14)\times 10^{-12} $  \\
 $ B $ (Cackett et al. 2013, $ \lambda = 4392\, \AA $) & $ (1.05 \pm 0.09)\times 10^{-12} $  \\
\hline                                   

\end{tabular}
\end{table}

As shown in Fig. \ref{sed}, the SED is consistent with a single black body contribution, likely due to the irradiated companion star (a similar result was obtained through the study of the correlations between UV and X-ray variability; Bernardini et al. 2013). No IR excess has been observed. In particular, a power-law fit ($ I_{\nu}\propto \nu^{\alpha} $) of the lower frequencies data reported an index $\alpha$ of $1.81\pm 0.14$, that is exactly opposite to what expected for an IR excess ($ \alpha\leq 0 $), and is on the contrary typical of a low frequency black body approximation ($ I_{\nu}\propto \nu^{2} $). In particular, the fit of the SED with a black body with fixed radius (corresponding to the Roche lobe dimension, $ 0.6\, R_{\odot} $) reported a black body temperature of $4050 \pm 30$ K, that is consistent with a K-spectral type main sequence star, and an irradiation luminosity of $ 4.5\,(\pm 5.5)\times 10^{32} \rm erg/s$, consistent with the X-ray luminosity reported in Campana et al. (2004).
The non-detection of a IR excess can not however totally exclude the emission of a jet, that could theoretically exist in Cen-X4 in quiescence, but it may only be detectable at radio frequencies.

If we hypothesize the effective emission of a jet from Cen X-4, we should expect at most a polarisation degree of a few $ \% $ in the optical (for instance 5$ \% $), that is typical of X-ray binaries jets with tangled magnetic fields. In order to obtain such a polarisation degree, we thus considered the most constraining 3$\sigma  $ upper limit we obtained in our analysis (that is $P < 0.5 \% $ in the $ I $-band) and evaluated the maximum possible contribution of the relativistic jet to the total $ I $-band flux to be at most the 10$ \% $. This upper limit to the jet emission is fairly constraining and is totally in agreement with the blue $ IR $ SED shown in Fig. \ref{sed}. 

Relativistic particles jets have been detected in some persistent neutron star X-ray binaries in IR and radio frequencies (i.e. Migliari et al. 2006), but there has not been ever a detection during quiescence. In particular, Migliari et al. 2006 predicted a steeper radio/X-ray correlation in neutron star X-ray binaries compared to black hole X-ray binaries, that would imply the possible emission of really faint jets from quiescent neutron stars, in accordance with our result.

\subsection{Thomson scattering}
\begin{figure}
\centering
\includegraphics[scale=0.25]{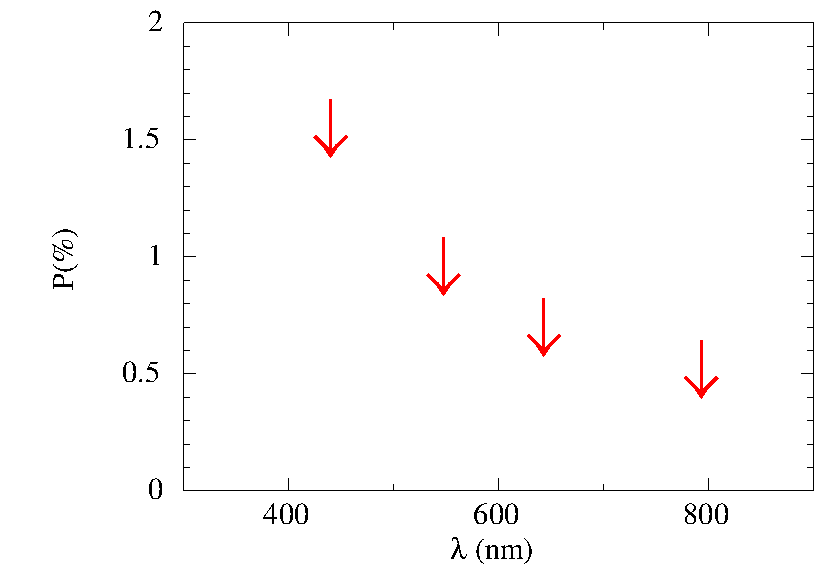}
\caption{Polarisation degree $ 3\sigma $ upper limits trend with respect to the wavelength of the four optical filters analysed, $ BVRI $.}
\label{sed_P}
\end{figure}

As explained in Sec. 1, scattering of radiation with electrons in the ionised disc can result in linear polarisation. The polarisation degree $ P $ that derives from this mechanism is expected to increase as a function of decreasing wavelength and not to be more then a few $ \% $ (Brown et al. 1978; Dolan 1984). In our case, the measured average polarisation degrees in the four filters never exceed the $ 1\% $ level, and for this reason could be consistent with the low polarisation degree expected in case of Thomson scattering with the free electrons in the disc. Unfortunately the low S/N ratio of our measures do not allow us to verify the expected increasing trend of $ P $ with decreasing wavelength (Fig. \ref{sed_P}). 

Furthermore scattering should also produce a linear polarisation degree that varies on time-scales of the orbital period (Kemp \& Barbour 1983; Dolan \& Tapia 1989; Gliozzi et al. 1998). 
In our case, an almost constant trend of the polarisation degrees in the four analysed filters has been observed. The only epoch that possesses a polarisation degree significantly higher then the average is in the $ V $-band, at phase 0.1. In this epoch, the companion star's contribution is at its minimum, i.e. the accretion disc should be mostly responsible of the observed variation. This fact grows in interest in light of the Thomson scattering hypothesis. Neverthless, the absence of similar behaviours in consecutive bands (and time), that in contrary showed almost constant $ P $, let this observation become less intriguing. 
We conclude that if there was a variation of $ P $ with the system's orbital phase, we were not able to detect it, due to the large error bars, often comparable to the possible obtained oscillation's semi-amplitudes.
\section{Conclusions}
In this work, we presented the results of an optical ($ BVRI $) and infrared ($ J $) polarimetric study on the LMXB Cen X-4 during quiescence, based on observations obtained in 2008 and 2007, respectively. 

We were searching for an intrinsic component of linear polarisation in the optical and IR for this source. 
We obtained a low polarisation degree $3\sigma$ upper limit in each optical filter, with the highest value in the $ B $-band ($P_{B}\leq 1.46 \% $) and lowest in the $ I $-band, where $ P $ is consistent with 0 within $ 1\sigma $. A $3\sigma$ upper limit to the linear polarisation of the $ 6 \% $ is obtained in the $ J $-band.

We built the SED of Cen X-4 from literature data, observing that it can be fitted by the only black body of the irradiated companion star. Assuming a typical expected polarisation degree of at most the $5\%$ for a NS X-ray binary jet with tangled magnetic field, our $ I $-band upper limit on the linear polarisation implies that the contribution from a possible jet should be relatively low ($ \lesssim 10\% $) in terms of flux. This is in agreement with the non-detection of an infrared excess in the SED of Cen X-4.

No variations correlated with the orbital period of the system has been detected and, due to the large error bars caused by the low S/N ratio, it was not possible to be conclusive on a possible increasing trend of $ P $ with decreasing wavelength, which should support the possibility of Thomson scattering of the radiation with accretion disc particles.
Observations with larger diameter telescopes (8 m class) would provide smaller uncertainties for the Stokes parameters (in particular in the $ B $-band, where Cen X-4 is fainter). This will be crucial to best investigate possible phase-orbital correlated variations of $ P $ and its wavelength trend in order to assess the origin of polarisation for Cen X-4.

\begin{acknowledgements}
We thank the anonymous referee for useful comments and suggestions.
MCB aknowledges P. M. Pizzochero for supportive
discussions and the INAF-Osservatorio Astronomico di Brera for kind hospitality during her master thesis. PDA acknowledges L. Monaco, E. Matamoros and A. Ederoclite for
their support during his observing run in La Silla. PDA acknowledges V. Lorenzi and
the TNG staff for their support in carrying out the NICS observations in service
mode.
\end{acknowledgements}



\begin{thebibliography}{52}
\expandafter\ifx\csname natexlab\endcsname\relax\def\natexlab#1{#1}\fi

\bibitem[{{Baglio} {et~al.}(2013){Baglio}, {D'Avanzo}, {Mu{\~n}oz-Darias},
  {Breton}, \& {Campana}}]{Baglio2013}
{Baglio}, M.~C., {D'Avanzo}, P., {Mu{\~n}oz-Darias}, T., {Breton}, R.~P., \&
  {Campana}, S. 2013, \aap, 559, A42

\bibitem[{{Bernardini} {et~al.}(2013){Bernardini}, {Cackett}, {Brown},
  {D'Angelo}, {Degenaar}, {Miller}, {Reynolds}, \& {Wijnands}}]{Bernardini2013}
{Bernardini}, F., {Cackett}, E.~M., {Brown}, E.~F., {et~al.} 2013, \mnras, 436,
  2465

\bibitem[{{Brocksopp} {et~al.}(2007){Brocksopp}, {Miller-Jones}, {Fender}, \&
  {Stappers}}]{Brocksopp07}
{Brocksopp}, C., {Miller-Jones}, J.~C.~A., {Fender}, R.~P., \& {Stappers},
  B.~W. 2007, \mnras, 378, 1111

\bibitem[{{Brown} {et~al.}(1978){Brown}, {McLean}, \& {Emslie}}]{Brown78}
{Brown}, J.~C., {McLean}, I.~S., \& {Emslie}, A.~G. 1978, \aap, 68, 415

\bibitem[{{Cackett} {et~al.}(2013){Cackett}, {Brown}, {Degenaar}, {Miller},
  {Reynolds}, \& {Wijnands}}]{Cackett2013}
{Cackett}, E.~M., {Brown}, E.~F., {Degenaar}, N., {et~al.} 2013, \mnras, 433,
  1362

\bibitem[{{Campana} {et~al.}(2004){Campana}, {Israel}, {Stella}, {Gastaldello},
  \& {Mereghetti}}]{Campana04}
{Campana}, S., {Israel}, G.~L., {Stella}, L., {Gastaldello}, F., \&
  {Mereghetti}, S. 2004, \apj, 601, 474

\bibitem[{{Canizares} {et~al.}(1980){Canizares}, {McClintock}, \&
  {Grindlay}}]{Canizares80}
{Canizares}, C.~R., {McClintock}, J.~E., \& {Grindlay}, J.~E. 1980, \apjl, 236,
  L55

\bibitem[{{Casares} {et~al.}(2007){Casares}, {Bonifacio}, {Gonz{\'a}lez
  Hern{\'a}ndez}, {Molaro}, \& {Zoccali}}]{Casares07}
{Casares}, J., {Bonifacio}, P., {Gonz{\'a}lez Hern{\'a}ndez}, J.~I., {Molaro},
  P., \& {Zoccali}, M. 2007, \aap, 470, 1033

\bibitem[{{Charles} {et~al.}(1980){Charles}, {Thorstensen}, {Bowyer}, {Clark},
  {Li}, {van Paradijs}, {Remillard}, {Holt}, {Kaluzienski}, {Junkkarinen},
  {Puetter}, {Smith}, {Pollard}, {Sanford}, {Tapia}, \& {Vrba}}]{Charles80}
{Charles}, P.~A., {Thorstensen}, J.~R., {Bowyer}, S., {et~al.} 1980, \apj, 237,
  154

\bibitem[{{Cheng} {et~al.}(1988){Cheng}, {Shields}, {Lin}, \&
  {Pringle}}]{Cheng88}
{Cheng}, F.~H., {Shields}, G.~A., {Lin}, D.~N.~C., \& {Pringle}, J.~E. 1988,
  \apj, 328, 223

\bibitem[{{Chevalier} {et~al.}(1989){Chevalier}, {Ilovaisky}, {van Paradijs},
  {Pedersen}, \& {van der Klis}}]{Chevalier89}
{Chevalier}, C., {Ilovaisky}, S.~A., {van Paradijs}, J., {Pedersen}, H., \&
  {van der Klis}, M. 1989, \aap, 210, 114

\bibitem[{{Conner} {et~al.}(1969){Conner}, {Evans}, \& {Belian}}]{Conner69}
{Conner}, J.~P., {Evans}, W.~D., \& {Belian}, R.~D. 1969, \apjl, 157, L157

\bibitem[{{Corbel} \& {Fender}(2002)}]{Corbel02}
{Corbel}, S. \& {Fender}, R.~P. 2002, \apjl, 573, L35

\bibitem[{{Covino} {et~al.}(1999){Covino}, {Lazzati}, {Ghisellini}, {Saracco},
  {Campana}, {Chincarini}, {di Serego}, {Cimatti}, {Vanzi}, {Pasquini},
  {Haardt}, {Israel}, {Stella}, \& {Vietri}}]{Covino1999}
{Covino}, S., {Lazzati}, D., {Ghisellini}, G., {et~al.} 1999, \aap, 348, L1

\bibitem[{{Cowley} {et~al.}(1988){Cowley}, {Hutchings}, {Schmidtke},
  {Hartwick}, {Crampton}, \& {Thompson}}]{Cowley88}
{Cowley}, A.~P., {Hutchings}, J.~B., {Schmidtke}, P.~C., {et~al.} 1988, \aj,
  95, 1231

\bibitem[{{D'Avanzo} {et~al.}(2005){D'Avanzo}, {Campana}, {Casares}, {Israel},
  {Covino}, {Charles}, \& {Stella}}]{Davanzo05}
{D'Avanzo}, P., {Campana}, S., {Casares}, J., {et~al.} 2005, \aap, 444, 905

\bibitem[{{di Serego Alighieri}(1998)}]{Serego}
{di Serego Alighieri}, S. 1998, Cambridge University Press, 199, 287

\bibitem[{{Dolan}(1984)}]{Dolan84}
{Dolan}, J.~F. 1984, \aap, 138, 1

\bibitem[{{Dolan} \& {Tapia}(1989)}]{Dolan89}
{Dolan}, J.~F. \& {Tapia}, S. 1989, \pasp, 101, 1135

\bibitem[{{Dubus} \& {Chaty}(2006)}]{Dubus2006}
{Dubus}, G. \& {Chaty}, S. 2006, \aap, 458, 591

\bibitem[{{Fender}(2001{\natexlab{a}})}]{FenderReview}
{Fender}, R. 2001{\natexlab{a}}, in The Second National Conference on
  Astrophysics of Compact Objects, 14

\bibitem[{{Fender}(2001{\natexlab{b}})}]{Fender01}
{Fender}, R.~P. 2001{\natexlab{b}}, \mnras, 322, 31

\bibitem[{{Gallo} {et~al.}(2007){Gallo}, {Migliari}, {Markoff}, {Tomsick},
  {Bailyn}, {Berta}, {Fender}, \& {Miller-Jones}}]{Gallo07}
{Gallo}, E., {Migliari}, S., {Markoff}, S., {et~al.} 2007, \apj, 670, 600

\bibitem[{{Gandhi} {et~al.}(2011){Gandhi}, {Blain}, {Russell}, {Casella},
  {Malzac}, {Corbel}, {D'Avanzo}, {Lewis}, {Markoff}, {Cadolle Bel}, {Goldoni},
  {Wachter}, {Khangulyan}, \& {Mainzer}}]{Gandhi11}
{Gandhi}, P., {Blain}, A.~W., {Russell}, D.~M., {et~al.} 2011, \apjl, 740, L13

\bibitem[{{Gliozzi} {et~al.}(1998){Gliozzi}, {Bodo}, {Ghisellini}, {Scaltriti},
  \& {Trussoni}}]{Gliozzi98}
{Gliozzi}, M., {Bodo}, G., {Ghisellini}, G., {Scaltriti}, F., \& {Trussoni}, E.
  1998, \aap, 337, L39

\bibitem[{{Haardt}(1993)}]{Haardt93}
{Haardt}, F. 1993, \apj, 413, 680

\bibitem[{{Hannikainen} {et~al.}(2000){Hannikainen}, {Hunstead},
  {Campbell-Wilson}, {Wu}, {McKay}, {Smits}, \& {Sault}}]{Hannikainen00}
{Hannikainen}, D.~C., {Hunstead}, R.~W., {Campbell-Wilson}, D., {et~al.} 2000,
  \apj, 540, 521

\bibitem[{{Helmboldt} {et~al.}(2007){Helmboldt}, {Taylor}, {Tremblay},
  {Fassnacht}, {Walker}, {Myers}, {Sjouwerman}, {Pearson}, {Readhead},
  {Weintraub}, {Gehrels}, {Romani}, {Healey}, {Michelson}, {Blandford}, \&
  {Cotter}}]{Helmboldt07}
{Helmboldt}, J.~F., {Taylor}, G.~B., {Tremblay}, S., {et~al.} 2007, \apj, 658,
  203

\bibitem[{{Hjellming}(1979)}]{Hjellming79}
{Hjellming}, R.~M. 1979, \iaucirc, 3369, 1

\bibitem[{{Kaluzienski} {et~al.}(1980){Kaluzienski}, {Holt}, \&
  {Swank}}]{Kaluzienski80}
{Kaluzienski}, L.~J., {Holt}, S.~S., \& {Swank}, J.~H. 1980, \apj, 241, 779

\bibitem[{{Kemp} \& {Barbour}(1983)}]{Kemp83}
{Kemp}, J.~C. \& {Barbour}, M.~S. 1983, \apj, 264, 237

\bibitem[{{McClintock} \& {Remillard}(1990)}]{McClintock90}
{McClintock}, J.~E. \& {Remillard}, R.~A. 1990, \apj, 350, 386

\bibitem[{{Migliari} {et~al.}(2006){Migliari}, {Tomsick}, {Maccarone}, {Gallo},
  {Fender}, {Nelemans}, \& {Russell}}]{Migliari2006}
{Migliari}, S., {Tomsick}, J.~A., {Maccarone}, T.~J., {et~al.} 2006, \apjl,
  643, L41

\bibitem[{{Migliari} {et~al.}(2007){Migliari}, {Tomsick}, {Markoff}, {Kalemci},
  {Bailyn}, {Buxton}, {Corbel}, {Fender}, \& {Kaaret}}]{Migliari07}
{Migliari}, S., {Tomsick}, J.~A., {Markoff}, S., {et~al.} 2007, \apj, 670, 610

\bibitem[{{Migliari} {et~al.}(2010){Migliari}, {Tomsick}, {Miller-Jones},
  {Heinz}, {Hynes}, {Fender}, {Gallo}, {Jonker}, \& {Maccarone}}]{Migliari10}
{Migliari}, S., {Tomsick}, J.~A., {Miller-Jones}, J.~C.~A., {et~al.} 2010,
  \apj, 710, 117

\bibitem[{{O'Dea}(1998)}]{Odea1998}
{O'Dea}, C.~P. 1998, \pasp, 110, 493

\bibitem[{{Pe'er} \& {Casella}(2009)}]{Peer09}
{Pe'er}, A. \& {Casella}, P. 2009, \apj, 699, 1919

\bibitem[{{Perlman} {et~al.}(2011){Perlman}, {Adams}, {Cara}, {Bourque},
  {Harris}, {Madrid}, {Simons}, {Clausen-Brown}, {Cheung}, {Stawarz},
  {Georganopoulos}, {Sparks}, \& {Biretta}}]{Perlman11}
{Perlman}, E.~S., {Adams}, S.~C., {Cara}, M., {et~al.} 2011, \apj, 743, 119

\bibitem[{{Rahoui} {et~al.}(2011){Rahoui}, {Lee}, {Heinz}, {Hines},
  {Pottschmidt}, {Wilms}, \& {Grinberg}}]{Rahoui11}
{Rahoui}, F., {Lee}, J.~C., {Heinz}, S., {et~al.} 2011, \apj, 736, 63

\bibitem[{{Russell} {et~al.}(2011){Russell}, {Casella}, {Fender}, {Soleri},
  {Pretorius}, {Lewis}, \& {van der Klis}}]{Russell11}
{Russell}, D.~M., {Casella}, P., {Fender}, R., {et~al.} 2011, ArXiv e-prints

\bibitem[{{Russell} \& {Fender}(2008)}]{Russell08}
{Russell}, D.~M. \& {Fender}, R.~P. 2008, \mnras, 387, 713

\bibitem[{{Russell} {et~al.}(2006){Russell}, {Fender}, {Hynes}, {Brocksopp},
  {Homan}, {Jonker}, \& {Buxton}}]{Russell06}
{Russell}, D.~M., {Fender}, R.~P., {Hynes}, R.~I., {et~al.} 2006, \mnras, 371,
  1334

\bibitem[{{Russell} {et~al.}(2007){Russell}, {Fender}, \& {Jonker}}]{Russell07}
{Russell}, D.~M., {Fender}, R.~P., \& {Jonker}, P.~G. 2007, \mnras, 379, 1108

\bibitem[{{Russell} \& {Shahbaz}(2014)}]{Russell14}
{Russell}, D.~M. \& {Shahbaz}, T. 2014, \mnras, 438, 2083

\bibitem[{{Schultz} {et~al.}(2004){Schultz}, {Hakala}, \&
  {Huovelin}}]{Schultz04}
{Schultz}, J., {Hakala}, P., \& {Huovelin}, J. 2004, Baltic Astronomy, 13, 581

\bibitem[{{Serkowski} {et~al.}(1975){Serkowski}, {Mathewson}, \&
  {Ford}}]{Serkowski75}
{Serkowski}, K., {Mathewson}, D.~S., \& {Ford}, V.~L. 1975, \apj, 196, 261

\bibitem[{{Shahbaz} {et~al.}(2008){Shahbaz}, {Fender}, {Watson}, \&
  {O'Brien}}]{Shahbaz08}
{Shahbaz}, T., {Fender}, R.~P., {Watson}, C.~A., \& {O'Brien}, K. 2008, \apj,
  672, 510

\bibitem[{{Shahbaz} {et~al.}(1993){Shahbaz}, {Naylor}, \&
  {Charles}}]{Shahbaz93}
{Shahbaz}, T., {Naylor}, T., \& {Charles}, P.~A. 1993, \mnras, 265, 655

\bibitem[{{Shahbaz} {et~al.}(2013){Shahbaz}, {Russell}, {Zurita}, {Casares},
  {Corral-Santana}, {Dhillon}, \& {Marsh}}]{Shahbaz13}
{Shahbaz}, T., {Russell}, D.~M., {Zurita}, C., {et~al.} 2013, \mnras, 434, 2696

\bibitem[{{Stetson}(1987)}]{Stetson1987}
{Stetson}, P.~B. 1987, \pasp, 99, 191

\bibitem[{{Torres} {et~al.}(2002){Torres}, {Casares}, {Mart{\'{\i}}nez-Pais},
  \& {Charles}}]{Torres02}
{Torres}, M.~A.~P., {Casares}, J., {Mart{\'{\i}}nez-Pais}, I.~G., \& {Charles},
  P.~A. 2002, \mnras, 334, 233

\bibitem[{{Wardle} \& {Kronberg}(1974)}]{Wardle1974}
{Wardle}, J.~F.~C. \& {Kronberg}, P.~P. 1974, \apj, 194, 249

\end{thebibliography}

\end{document}